
%
%
%
\input phyzzx
\newif\iffiginclude
\figincludefalse        
\vsize=54pc
\unnumberedchapters
\date={September 1993}
\Pubnum={\caps UPR-584-T
 }
\iffiginclude
  \input psfig
\fi

\def\to{\rightarrow}
\titlepage
\title{Review of $Z'$ Physics at Future Colliders\foot{Based on
talks presented by M. Cveti\v c at the
Workshop on Physics and Experiments at Linear $e^+e^-$ Colliders, Waikoloa,
Hawaii, April 26-30, 1993 and the International Europhysics
 Conference on High Energy
Physics, Marseilles, July  21-28, 1993.}}
\frontpageskip=0.5\medskipamount plus 0.5 fil
\author{M. Cveti\v c$^*$, F. del Aguila$^+$  and P. Langacker$^*$}
\address{*Department of Physics,\break
University of Pennsylvania,
Philadelphia, PA 19104--6396\break
 +Departamento de F\'\i sica Te\'orica y del Cosmos,\break
Universidad de Granada,
Granada, 18071, Spain\break }

\singlespace
 \abstract{We present recent developments in the diagnostic study of
heavy gauge bosons at future colliders  with the  emphasis  on the
determination of gauge couplings of $Z'$ to quarks and leptons.
Proposed  diagnostic probes at
 future hadron colliders  are discussed. The study of
 statistical uncertainties expected for the above  probes
at  the LHC  and $M_{Z'}\simeq 1$ TeV shows that three out of four normalized
couplings can be determined,   thus  allowing for a  distinction between
models. The complementary nature of the  probes at future $e^+e^-$
colliders is also discussed.}

 \REF\ACLII{F. del Aguila, M. Cveti\v c, and P. Langacker,
 in these Proceedings.}
 \REF\LRR{ P.~Langacker, R.~Robinett, and J.~Rosner, Phys. Rev.
D {\bf30}, 1470 (1984).}
 \REF\Bar{ V.~Barger, {\it et al.}, Phys. Rev.   {\bf D 35}, 2893
(1987).}
\REF\DuLa{ L.~Durkin and P.~Langacker, Phys. Lett. {\bf B166},
436 (1986); F.~del~Aguila, M.~Quiros, and F.~Zwirner, Nucl. Phys. B
{\bf287}, 419 (1987); {\bf284}, 530 (1987); J.~Hewett and T.~Rizzo in
{\sl Proceedings of the 1988 Snowmass Summer Study on High Energy
Physics in the 1990's}, Snowmass, CO 1988; P.~Chiappetta  {\it et
al.}, in the {\sl Proceedings of the Large Hadron Collider Workshop},
Aachen, Germany, 1990; J. Hewett and T. Rizzo, Phys. Rep. {\bf 183}, 193
(1989.}
\REF\HRII{J. ~Hewett and  T. ~Rizzo, MAD/PH/649/91.}

{\bf\noindent 1. Introduction}
\vglue 0.4cm
Extended gauge structures, including  heavy neutral ($Z'$) and
charged ($W'$)
 gauge bosons, are an essential part of theories beyond the
standard models, including grand unified theories
(GUT) and superstring theory.
 If heavy gauge bosons
turn out to have a mass in the few TeV region, future hadron colliders,
the Superconducting Super Collider (SSC) and the Large Hadron Collider (LHC),
  would be an ideal place to discover them.\foot{For a review of
present bounds on $Z'$ see for example Ref. \ACLII .}

 At such hadron colliders heavy neutral  gauge bosons $Z'$
 with  mass  up to  around 5 TeV  can be produced and
clearly detected via leptonic two fermion  decays   $pp \to Z'\to \ell^+
\ell^-$ ($\ell=e,\mu$).\refmark{\LRR -\HRII}\
 After the discovery of a new gauge boson  one would like to
learn more about its  properties.
There would be a need to address and possibly  separate different features:
{\it (i)} $Z'$ couplings to ordinary
fermions,{\it  (ii)} the nature of symmetry breaking,
and {\it (iii)} $Z'$ couplings to exotic
fermions and supersymmetric partners.
\REF\ZWIRNER{
F.~del Aguila, M.~Quir\'os, and F.~Zwirner, Nucl. Phys.
B {\bf284}, 530 (1987); P.~Kalyniak and M.~Sundaresan, Phys. Rev. D
{\bf35}, 75 (1987).}\
\REF\CKL{M.~Cveti\v c,  P.~Langacker, and B.~Kayser,
Phys. Rev. Lett. {\bf 68}, 2871 (1992).}
\REF\GUNION{
N.~Deshpande, J.~Gunion, and F.~Zwirner, in the
{\sl Proceedings of the Workshop on Experiments Detectors and
Experimental Areas for the Supercollider} (Berkeley 1987); N.
Deshpande, J.~Grifols, and A.~M\'endez, Phys. Lett. {\bf B208}, 141
(1988).}\REF\AGUILA{
F.~del Aguila, L.~Ametller, R.~Field, and
L.~Garrido, Phys. Lett. {\bf B201}, 375 (1988); Phys. Lett. {\bf
B221}, 408 (1989).}

A test of the  symmetry breaking structures are the decays $Z'\to
W^+W^-$\refmark{\ZWIRNER ,\CKL},  which are suppressed by $Z-Z'$ mixing but
still  have a sizable rate due to the
enhancement of  the longitudinal components of the $W$ bosons.
However, they suffer from serious QCD backgrounds.\refmark{\GUNION,\AGUILA}
In  theories with  charged gauge bosons, {\it e.g.}, left-right (LR)
symmetric models, the ratio $M_{Z'}/M_{W'}$ plays an  analogous
role to
the $M_Z/M_W$  ratio (related
to the  $\rho$ parameter)   in the standard model.
 This ratio
therefore  yields indirect  information on the nature of the
Higgs sector.\Ref\CLO{M. Cveti\v c and
P. Langacker, Phys. Rev. {\bf D42}, 1797 (1990).}

The study of $Z'$ decays into exotic
particles  also  yields  useful information.
In particular, $W'\to \ell N$ and $Z'\to NN$ and subsequent  decays of
heavy right-handed neutrinos  $N$ turn out
to be useful probes for distinguishing
the left-right models from those with only an additional $U(1)$.\Ref\DL{M. J.
Duncan and P. Langacker, Nucl. Phys. {\bf B277}, 285 (1986).
}

 In the following we shall concentrate on the diagnostic study of $Z'$
couplings to quarks and leptons.  In Section 2 we address the corresponding
 probes at future hadron colliders. In Section 3 we present an
analysis for extracting information about the couplings. In Section 4
a comparison with   $e^+e^-$ colliders
is given. Conclusions are given in Section 5.
\vglue 0.6cm
{\bf\noindent 2. Probes for ${\bf Z'}$ gauge couplings at hadron colliders}
\vglue 0.4cm

In the main production channels,  $pp \to Z'\to \ell^+
\ell^-$ ($\ell=e,\mu$), one would be able to measure immediately
the mass $M_{Z'}$, the width $\Gamma_{tot}$ and the total cross section
 $\sigma_{\ell \ell}$. However, $\sigma_{\ell\ell}=\sigma (pp\to Z')
 B$  is {\it not}   a useful diagnostic
probe for the $Z'$ gauge couplings to quarks and leptons.
While $\sigma(pp\to Z')$,   the total  production cross section, can be
calculated to within  a few percent for
given $Z'$ couplings, the branching ratio into leptons,
$B\equiv\Gamma(Z'\to\ell^+\ell^-)/\Gamma_{tot}$,
is model dependent; it depends on the contribution of exotic
fermions and supersymmetric partners to the
$Z'$ width, and thus it  cannot be
useful as a diagnostic test for the $Z'$ gauge couplings.  However, it
would be a useful indirect probe for the existence of the exotic
fermions or superpartners.
On the other hand, from measurements of the total width\refmark\HRII\
$\Gamma_{tot}$,  and $\sigma_{\ell \ell}$ one obtains
$\sigma \Gamma(Z'\rightarrow \ell^+\ell^-)\equiv\sigma B\Gamma_{tot}$, which
 probes the absolute magnitude of the gauge couplings.
 \REF\AV{F. del Aguila and J. Vidal,
Int. J. of Math. Phys. {\bf A4}, 4097 (1989).}
\REF\CLII{M.~Cveti\v c and P. Langacker,
 Phys. Rev. {\bf D 46}, R14 (1992).}
\REF\AAC{J. Anderson, M. Austern, and B. Cahn,  Phys.
Rev. Lett. {\bf 69}, 25 (1992) and Phys. Rev. {\bf D46}, 290 (1992).}
\REF\FT{ A. Fiandrino and  P. Taxil, Phys. Rev. {\bf D44}, 3409 (1991);
 Phys. Lett.
{\bf B292}, 242 (1992).}
\REF\PACO{F. del Aguila, B. Alles, L. Ametller and
A. Grau, Phys. Rev. {\bf D48}, 425 (1993).}
\REF\HR{J. Hewett and T. Rizzo, Phys. Rev. {\bf D47}, 4891 (1993).}
\REF\CLIV{M.~Cveti\v c and P. Langacker,  Phys. Rev. {\bf D46}, 4943 (1992).}
\REF\RII{T. Rizzo, Phys. Rev. {\bf D47}, 965 (1993).}
\REF\ACL{F. del Aguila, M.
Cveti\v c and P. Langacker, Phys. Rev. {\bf D48}, R969
(1993).}
\REF\RM{A. Henriques and L. Poggioli, ATLAS
Collaboration, Note PHYS-NO-010 (October 1992);
T. Rizzo, ANL-HEP-PR-93-18 (March 1993).}
\REF\MOH{P. Mohapatra,  Mod. Phys. Lett. {\bf A8}, 771 (1993) and these
Proceedings.}
\REF\HRIII{J. Hewett and T. Rizzo, ANL-HEP-CP-93-51 (August 1993).}

In the following we
will  address signals  which probe {\it  relative strengths} of $Z'$ gauge
couplings. Until recently, the only recognized
 probe for the  gauge couplings at future hadron  colliders was  the
 forward-backward asymmetry\refmark\LRR\ in
the main  production channel  $pp\to Z'\to
\ell^+\ell^- \ (\ell=e$ or $\mu$).\foot{See also Ref. \AV . }\
There was a clear
 need to search for  additional, complementary  probes, thus  triggering
 a   renewed interest\refmark{\CKL , \CLII -\RM } in diagnostic probes
of the $Z'$ couplings.
 The previously  poorly  explored domain of $Z'$ diagnostics  is
 now a rich and ongoing field.
\REF\RI{T. Rizzo, Phys. Lett.  {\bf B192}, 125 (1987).}

The nature of such probes can be classified according to the
type of  channel in which they can be measured:
\bigskip
{\noindent (Ia) {\it The main  production channels}:}

{ (A) Forward-backward asymmetry,\refmark{\LRR}}

{ (B) Ratio of cross-sections in different rapidity bins,\refmark\ACL }

{ (C) Corresponding asymmetries
if  proton polarization  were available.\refmark\FT}
\smallskip
{\noindent (Ib) {\it  Other two-fermion  final state  channels}:}

{ (D) Measurements of the $\tau$ polarization
 in the $pp\to Z'\to  \tau^+\tau^-$
channel,\refmark\AAC }

{ (E) Measurements of the cross section in the $pp\to Z'\to   jet\  jet$
channel.\refmark{\RM, \MOH} }
\smallskip
{\noindent (II) {\it  The four-fermion final state  channels}:}

{ (F) Rare decays $Z'\to W\ell\nu_{\ell}$,\refmark{\RI, \CLII}}

{ (G) Associated productions
$pp\rightarrow Z' V$
with $V=(Z,W)$\refmark\CLIV\  and $V=\gamma$\refmark \RII .}

Probes under  (Ia) constitute   distributions, {\it i.e.},
``refinements'', in the main production channels.
The  forward-backward asymmetry  (A) was long recognized as useful to
probe a particular combination of
quark and lepton couplings. On the other hand,
the rapidity ratio (B)\refmark\ACL\  was recognized as
a useful complementary probe separating the $Z'$ couplings to
the $u$ and $d$ quarks due to the  harder valence $u$-quark distribution
in the proton relative to the $d$-quark.
Probes (C) are very useful if
 proton polarization were available at future hadron colliders.

 For probes in  other than  the main production channels (Ib)
the   background can be large. For  (E) the QCD background may
be difficult to overcome.\refmark{\RM , \MOH}
(D)  provides another interesting possibility to address the $Z'$  lepton
couplings, while (E) is the only  probe available for the left-handed quark
coupling.\refmark\CLIV

Probes in the  four-fermion final state  channels (II)
have suppressed rates  compared  to the two fermion channels (Ia) and (Ib).
 In these cases one hopes to  have enough statistics, and no attempt to study
distributions  seems to be possible.

Rare decays   $Z'\rightarrow f_1\bar f_2 V$,
with  ordinary  gauge bosons $V=(Z,W)$ emitted
by Brems--strahlung from one of the
fermionic  ($f_{1,2}$) legs turn out to have sizable statistics, \refmark{\RI
,\CLII}\ which
is  due to a  double  logarithmic
enhancement,\refmark{\CLII}\  closely related
to collinear and infrared singularities of gauge theories. They
 were studied in detail in Refs. {\CLII, \PACO , \HR }.
A background study\refmark{\CLII, \CLIV} of
such  decays  revealed that the only useful  mode\foot{$Z'\to Z \ell^+\ell^-$
does not significantly discriminate between models.}  without
large standard model and QCD backgrounds
is (F): $Z'\to W\ell \nu_{\ell}$  and $W\to hadrons$,
with the imposed  cut $m_{T\ell\nu_\ell}>90$ GeV
on the transverse mass of the
$\ell \nu_\ell$.
(This assumes that there is a sufficiently high efficiency for the
reconstruction of $W\to hadrons$ in events tagged by an energetic lepton.
Further study of the QCD background and the jet reconstruction for such
processes is needed.)
 The same mode   with
$W\to \ell\nu_\ell$ may also   be detectable\refmark\PACO\  if appropriate
cuts are applied.\foot{ A  (remote) possibility
of gaining useful information from
$Z'\to Z\nu_{\ell}\bar\nu_{\ell}$\refmark{\CLII , \CLIV ,\HR}\
 was  recently resurrected.\refmark{\HRIII}}
 These modes probe a left-handed leptonic coupling.

 Associated productions  (F) turn out to be relatively clean
signals\refmark\CLIV\  with slightly smaller statistics than rare decays.
They probe  a particular combination  the gauge couplings to quarks
and are thus complementary  to  rare decays.

 At  the SSC and the LHC the above signals are feasible
 diagnostic probes  for $M_{Z'}\lsim 1-2$ TeV.
For diagnostic study of $Z'$ couplings large
luminosity is  important;  for
$M_{Z'}\sim 1$ TeV, one expects about
 twice  as many events at
 the LHC  (projected luminosity
$10^{34} \hbox{cm}^{-2}\hbox{s}^{-1} $) than at the SSC (projected luminosity
$10^{33} \hbox{cm}^{-2}\hbox{s}^{-1} $).
For higher $Z'$ masses the number of events drops rapidly. For
$M_{Z'}\simeq 2$ TeV, the statistical errors on forward-backward asymmetry
(A), the
rapidity ratio  (B), and rare decays (F)
 increase by a factor of  4, while those on associated productions (G)
 increase by a factor of 3.  A
reasonable discrimination between models and determination of
the couplings may still be  possible, primarily from the forward-backward
asymmetry and the rapidity ratio. However, for
$M_{Z'}\simeq 3$ TeV the statistical errors on the first three quantities
are larger  by a factor of 13 than for 1 TeV, and there are not enough
events expected for
the associated production to allow a meaningful measurement.
 For $M_{Z'}\gsim 3$ TeV, there is therefore little ability
to discriminate  between models.

\vglue 0.6cm
{\bf\noindent 3. Determination of gauge couplings  at hadron colliders}
\vglue 0.4cm
We would now like to  examine how well the various $Z'$
couplings could be extracted from  the above probes. We will
mainly concentrate on probes (A), (B), (F) and (G), which , from our
perspective,  are most feasible.
 For definiteness, we consider the statistical uncertainties for
  $M_{Z'}=$1 TeV at the LHC.
Eventually, the uncertainties associated with the detector acceptances
and  systematic errors will have to be taken into account.

We consider the following typical models:
$Z_\chi$  in $SO_{10}\rightarrow SU_5\times U_{1\chi}$,
$Z_\psi$ in $E_6\rightarrow SO_{10}\times U_{1\psi}$,
$Z_\eta=\sqrt{3/8}Z_\chi-\sqrt{5/8}Z_\psi$  in
superstring inspired models in which $E_6$ breaks directly to a rank 5
group,and
$Z_{LR}$ in LR symmetric models.  For conventions
in the neutral current interactions see
\REF\LL{P. Langacker and M. Luo, Phys. Rev. {\bf D 44},
 817 (1991).} Ref. \LL .
In the following we assume family universality,
 neglect $Z-Z'$ mixing  and  assume $[Q',T_i]=0$,
which holds for $SU_2 \times U_1 \times U_1'$ and LR models.
 Here,  $Q'$ is the $Z'$ charge  and $T_i$ are the $SU_{2L}$ generators.

\input tables
\pageinsert
\begintable
\| $\chi$\ \ \  & $\psi$\ \ \  & $\eta$\ \ \ & $LR$\ \ \ \crthick
$\gamma^\ell_L$ \| $0.9\pm 0.018$ & $0.5\pm 0.03$& $0.2\pm 0.015$ &
$0.36\pm 0.007$
\nr
$\gamma^q_L$ \| 0.1\ \ \  & 0.5\ \ \  & 0.8\ \ \  & 0.04\ \ \ \nr
$\tilde{U}$ \| $1\pm 0.18$ & $1\pm 0.27$ & $1\pm 0.14$ & $37\pm 8.3$ \nr
$\tilde{D}$ \| $9\pm 0.61$ & $1\pm 0.41$ & $0.25\pm 0.29$ & $65\pm 14$ \nr
$\rho_{ud}$\| --0.19& --0.24& --0.66&0.93\endtable
\vskip0.2in
\noindent{\bf Table I}\refmark\ACL\ Values of $\gamma_L^\ell$, $\gamma_L^q$,
 $\tilde{U}$,
and $\tilde{D}$
for the $\chi$, $\psi$, $\eta$, and ${LR}$
 models. The error bars indicate how well the coupling could be
measured at the LHC for $M_{Z'}=1$ TeV. $\rho_{ud}$ indicates the correlation
coefficient between $\tilde U$ and $\tilde D$.
Except for the $\chi$ model the correlation  between $\gamma_L^\ell$
and ($\tilde U, \  \tilde D$) are negligible. \hfil\break
\line{\hrulefill}
\vfill
\iffiginclude
\line{\hfil\psfig{figure=image.ps,height= 4.5in}\hfil}
\fi
\line{\hrulefill}
{\bf Figure 1} 90\%  confidence level ($\Delta \chi^2=6.3$)
contours   for the $\chi$, $\psi$ and
$\eta$ models are plotted for $\tilde U$, versus  $\tilde D$, versus
$\gamma_L^\ell$. The input data are for $M_{Z'}=1$ TeV  at the LHC
and include statistical errors only.  The axes $\tilde U$,  $\tilde D$, and
$\gamma_L^\ell$  are rescaled by a factor 20, 2, and 40, respectively.
\endinsert
The relevant quantities\refmark{\CLIV\ ,\ACL}\  to
 distinguish different theories are
the charges, $\hat{g}^u_{L2}=\hat{g}^d_{L2}\equiv\hat{g}^q_{L2}$,
$\hat{g}^u_{R2}$, $\hat{g}^d_{R2}$, $\hat{g}^\nu_{L2}=\hat{g}^e_{L2}
\equiv\hat{g}^\ell_{L2}$, and $\hat{g}^\ell_{R2}$, and the gauge
coupling strength $g_2$.
The signs of the charges will be hard to determine at hadron
colliders. Thus the following
 four ``normalized''
observables can be probed:\refmark\CLIV
\def\denom{{(\hat{g}^\ell_{L2})^2+(\hat{g}^\ell_{R2})^2}}
$$\gamma_L^\ell\equiv{{(\hat{g}^\ell_{L2})^2}\over\denom},\ \
\gamma_L^q\equiv{{(\hat{g}^q_{L2})^2}\over\denom},\ \
\tilde{U}\equiv{{(\hat{g}^u_{R2})^2}\over {(\hat{g}^q_{L2})^2}},\ \
\tilde{D}\equiv{{(\hat{g}^d_{R2})^2}\over {(\hat{g}^q_{L2})^2}}.\eqn\tild $$
The values of  $\gamma_L^\ell$, $\gamma_L^q$,
 $\tilde{U}$, and $\tilde{D}$  for the above models  are listed in
Table~I.

 The forward-backward asymmetry  (A) is defined as:
  $A_{FB}= {{\left[\int_{0}^{y_{max}}-\int_{-y_{max}}^0\right]
[F(y)-B(y)]dy}\over
 {\int_{-y_{max}}^{y_{max}} [F(y)+B(y)]dy}}, $  while the
 rapidity ratio (B) is defined as:\refmark\ACL\
$r_{y_1}= {{\int_{-y_1}^{y_1}[F(y)+B(y)]dy}
\over{(\int_{-y_{max}}^{-y_{1}}+\int_{y_1}^{y_{max}}) [F(y)+B(y)]dy}}$.
Here $F(y)\pm B(y)=[\int_0^1\pm\int_{-1}^0] \,d\cos\theta (d^2\sigma/
dy\, d\cos\theta)$, where  $y$  is the $Z'$ rapidity and $\theta$ is the
$\ell^-$ angle in the $Z'$ rest
frame. The rapidity range is  from $\{ -y_{max}, y_{max}\}$.
 $y_1$ is chosen in a range $0<y_1<y_{max}$ so that the
number of events in the two bins are comparable.
 At the LHC ($y_{max}\simeq 2.8$) for  $M_{Z'}\simeq 1$ TeV,  and $y_1=1$
 turns out to be an appropriate choice.
For rare decays (F)  one defines:\refmark\CLII\
$r_{\ell\nu W} \equiv {{B(Z'\rightarrow W\ell\nu_\ell)}
\over {B(Z'\rightarrow\ell^+\ell^-)}}$, in which one sums
over $\ell=e,\mu$ and  over $W^+$, $W^-$.
For the associated productions (G)
 one   defines\refmark\CLIV\
 the ratios:
$R_{Z'V}={{\sigma
(pp\to Z'V)B(Z'\to \ell^+\ell^-)}\over{
\sigma (pp\to Z')B(Z'\to \ell^+\ell^-)}}$
 with $V=(Z, W)$\   decaying into leptons   and  quarks, and $V=\gamma$\
with an imposed $p_{T\gamma}\geq 50$ GeV cut.
 $\ell$ includes both
$e$ and $\mu$.

Statistical   errors of the above probes
for $M_{Z'}=1$ TeV
at the  LHC  are given in Ref.  \ACL .
They turn out to be sufficiently small
 to distinguish  between models.
 The six quantities  $A_{FB}$,
$r_{y_1}$, $r_{\ell\nu W}$, and $R_{Z'V}$ with $(V=Z,W,\gamma)$
yield significant information on  three
($\gamma_L^\ell$, $\tilde U$ and $\tilde
D$) out of four normalized gauge couplings
 of ordinary fermions to the $Z'$.
The fourth normalized coupling    $\gamma_L^q$
could be determined\refmark{\CLIV}\ by a measurement of the branching ratio
$B(Z'\rightarrow q{\bar q})$ (E). Recent studies indicate that this might be
possible.\refmark{\RM ,\MOH }

To study  with what precision these couplings could be determined,
a  combined $\chi^2$ analysis of these observables has been
performed.\refmark\ACL\
Only the statistical uncertainties have been included and
 correlations between the observables have been ignored.
The results are  given in  Table I.
In particular, $\gamma_L^\ell$ can be determined very well (between
2\% and 8\% for the $\chi$, $\psi$, and $\eta$ models), primarily due to
the small statistical error for the rare decay mode $Z'\to W\ell\nu_\ell$.
On the other hand the  quark couplings have larger uncertainties,
typically 20\% for $\tilde U$, and an absolute error of $\sim 0.3 - 0.6$ for
$\tilde D$ (except $Z_{LR}$).
In  Figure 1 90\% confidence level  ($\Delta \chi^2=6.3$) contours\foot{
The 90\%
confidence level  contours for  projections
on the more familiar  two-dimensional
parameter subspaces correspond to $\Delta \chi^2=4.6$.}
are given in a three-dimensional plot for $\tilde U$ versus $\tilde D$
versus $\gamma_L^\ell$ for
the  $\eta$, $\psi$ and $\chi$ models. The $LR$ model
has  $\tilde U$ and $\tilde D$ in a
different region of the parameter space  (see
Table I).  From Figure 1 it is clear
that  one can distinguish well between  different  models.

We would also like to comment briefly on the determination of
heavy charged ($W'$)  gauge
boson  couplings. While the forward-backward asymmetry
in the main production channels $pp\to W'\to \ell\nu_\ell$  ($\ell=e,\mu $)
probes some  combination of gauge couplings, it does not distinguish $\hat
g_{L2}$
from $\hat g_{R2}$ couplings. On the
other hand  rare decays
$W^{\prime\pm}\to W\ell^+\ell^-$ and associated productions
$pp\to W^{\prime\pm}W^\mp $ are strongly
suppressed\refmark{\CLII ,\CKL ,\CLIV}
 if $W'$ has only $\hat g_{R2}$ couplings, as in the LR symmetric models.
In models where $W'$
has $\hat g_{L2}$ couplings, {\it e.g.},
the so-called un-unified models,\Ref\GJS{H. Georgi, E.
Jenkins and E. Simmons, Phys. Rev. Lett. {\bf 62}, 2789 (1989); {\it ibid}
{\bf 63}, 1540 (1989){\bf E} ; Nucl. Phys. {\bf B331}, 541 (1990).}\ the
corresponding rates are, however, not suppressed; primarily due to the
larger gauge couplings of $W'$ , the
corresponding  rates  allow for determination of $\hat g_{L2}$
couplings for $M_{W'}$
up to around 3 TeV.\Ref\CLL{M. Cveti\v c, P. Langacker
and J. Liu, Univ. of Pennsylvania preprint, UPR-0579-T (August
1993).}

\vglue 0.6cm
{\bf\noindent 4. Comparison with  $e^+e^-$ colliders }
\vglue 0.4cm
\REF\DLRSV{A. Douadi, A. Leike, T. Riemann, D. Schaile and C. Verzegnassi, Z.
Phys.  {\bf C56}, 289 (1992).}
\REF\HRV{J. Hewett and T. Rizzo, ANL-HEP-CP-91-90, 1991; T. Rizzo,
ANL-HEP-CP-91-96, 1991.}
Future $e^+e^-$ colliders (next linear collider
(NLC))  with large enough  center of mass energy, {\it
e.g.}, $\sqrt s=$ 2 TeV,  could provide a clean way to discover and study
the  properties of  a $Z'$.  A  more likely
possibility is a NLC with center of mass energy in the
range of
$\sqrt s=$ 500 GeV, which would, due to  interference effects of the  $Z'$
propagator with the $\gamma$ and $Z$, provide  complementary
 diagnostics  of the  $Z'$ .\refmark{\DLRSV , \HRV}\
In order to distinguish between models  the standard probes
$\sigma_{lep}$,
$A_{FB}^{lep}$ and $R^{had}={{\sigma^{had}}\over{\sigma^{lep}}}$,
as well as
$A_{LR}^{lep, had}$ (if the  $e$ polarization is available) can be used.
\refmark{\DLRSV ,\HRV }\
Such probes allow for distinguishing between models for $M_{Z'}$ for up to
around 1 TeV.  In such studies one has to take into account
 radiative QED corrections and the  experimental set up.\refmark\DLRSV\

When addressing  diagnostics  of $Z'$ gauge couplings\Ref\ACLII{F. del
Aguila, M. Cveti\v c and P. Langacker, in preparation.}   the above probes
are complementary  to the ones at future hadron colliders (as discussed  in
Chapters  2  and 3).
In particular,  since at the NLC  the effects of a  $Z'$ are due to the $Z'$
propagator  interference terms, the $Z'$ total width $\Gamma_{tot}$ is not
measurable,  while  the mass $M_{Z'}$  and the  absolute value of the gauge
coupling $g_2$  is measurable only in a combination, {\it e.g.},  ${{{g_2^2}
({\hat{g}^\ell_{V2}})^2}\over  {M_{Z'}^2}}$.
On the other hand,  the  four normalized    charges, {\it e.g.},
$\hat{g}^\ell_{A2}\over\hat{g}^\ell_{V2}$
$\hat{g}^u_{R2}\over \hat{g}^\ell_{V2}$,
$\hat{g}^d_{R2}\over\hat{g}^\ell_{V2}$,
and $\hat{g}^q_{L2}\over\hat{g}^\ell_{V2}$,\foot{
Since the $\ell$ couplings to
$Z$ have the property $\hat g_{L1}^\ell\simeq -\hat g_{R1}^\ell$
it turns out that
the above probes single out the $Z'$ leptonic couplings primarily in the
combinations $\hat g_{V2, A2}^\ell\equiv
\hat g_{L2}^\ell\pm\hat g_{R2}^\ell$.}\  (and not only the  corresponding
normalized {\it squares} of charges as in
 Eq.\tild ) can be probed.  $\sigma_{lep}$ and
 $A_{FB}^{lep}$ probe
 ${({\hat{g}^\ell_{A2}})^2\over ({\hat{g}^\ell_{V2}})^2}$,
 which is complementary to  $\gamma_L^\ell$.  On the
other  hand, $R^{had}$ provides
information on one   combination of normalized quark couplings.
 If the $e$  polarization  were available the
$A_{LR}^{lep}$ would be a  sensitive  probe for  ${{\hat{g}^\ell_{A2}}
\over {\hat{g}^\ell_{V2}}}$, while
$A_{LR}^{had}$ would  yield information on
another  combination of normalized quark couplings.
 The possibility of flavor tagging in
the heavy quark  ($c,b,t$) sector would allow for measuring   $R^{u,d}$ and
$A_{FB}^{u,d}$, which would in  turn  provide  complementary
information  on all the quark couplings.

\vglue 0.6cm
{\bf\noindent 5. Conclusions }
\vglue 0.4cm

If there are new heavy gauge bosons with $M_{Z'}\lsim 1-2$ TeV
 future hadron colliders  would not only be an ideal place for their
discovery, but would also provide a fertile ground to learn more about
their properties, in particular, their couplings to quarks and leptons.
Future
$e^+e^-$ colliders with $\sqrt s \sim 500 $ GeV would in
turn allow for complementary
studies of heavy gauge boson  properties.

{\bf Acknowledgment} This work was supported
by the Department of Energy Grant \#
DE--AC02--76--ERO--3071, and the
Texas National Research Commission Laboratory.
We would like to thank J. Liu  for enjoyable collaboration and D. Benton  for
 help with the Figure.

\refout\end